\documentclass[sigconf]{acmart}
\copyrightyear{2020}
\acmYear{2020}
\setcopyright{acmcopyright}
\acmConference[SIGIR '20]{Proceedings of the 43rd International ACM SIGIR Conference on Research and Development in Information Retrieval}{July 25--30, 2020}{Virtual Event, China}
\acmBooktitle{Proceedings of the 43rd International ACM SIGIR Conference on Research and Development in Information Retrieval (SIGIR '20), July 25--30, 2020, Virtual Event, China}
\acmPrice{15.00}
\acmDOI{10.1145/3397271.3401193}
\acmISBN{978-1-4503-8016-4/20/07}
\usepackage{graphicx}
\usepackage{subfigure}
\usepackage{bbm}
\usepackage{booktabs}
\usepackage{multirow}
\usepackage{diagbox}
\usepackage{algorithm}
\usepackage{algorithmic}
\usepackage{multirow}
\usepackage{xcolor}

\begin{document}
\fancyhead{} 

\title{DCDIR: A Deep Cross-Domain Recommendation System for Cold Start Users in Insurance Domain}
\author{Ye Bi, Liqiang Song, Mengqiu Yao, Zhenyu Wu, Jianming Wang, Jing Xiao}
\email{magicyebi@163.com, {songliqiang537, yaomengqiu621, wuzhenyu447, wangjianming888, xiaojing661}@pingan.com.cn}
\affiliation{%
\institution{Ping An Technology Shenzhen Co., Ltd}
}

\begin{abstract}
Internet insurance products are apparently different from traditional e-commerce goods for their complexity, low purchasing frequency, etc. So, cold start problem is even worse. In traditional e-commerce field, several cross-domain recommendation (CDR) methods have been studied to infer preferences of cold start users based on their preferences in other domains. However, these CDR methods couldn’t be applied into insurance domain directly due to product complexity. In this paper, we propose a Deep Cross-Domain Insurance Recommendation System (DCDIR) for cold start users. Specifically, we first learn more effective user and item latent features in both domains. In target domain, given the complexity of insurance products, we design a meta-path based method over insurance product knowledge graph. In source domain, we employ GRU to model users' dynamic interests.  Then we learn a feature mapping function by multi-layer perceptions . We apply DCDIR on our company’s dataset, and show DCDIR significantly outperforms the state-of-the-art solutions.
\end{abstract}



\maketitle

\section{Introduction}
 
Nowadays, internet finance is booming and rapidly infiltrating into all kinds of traditional financial fields. Internet insurance adapted to the trend of economic boom in internet age, since it can not only overcome the limitations of live sales and geography, but also provide savings for both companies and their consumers. Due to nature of insurance industry, the products that insurance companies can provide on internet always have the following characteristics: 1) the coverage time is no more than 1 year; 2) the prices are lower than long-term insurances; 3) they covers widely, including property and casualty, etc.; 4) the customers are not required to buy other insurance products earlier. However, recommending insurance products online is challenging. First, insurance policies are so complex that ordinary users are relatively lack of knowledge to understand them. Besides, insurance products are typically bought to be used for a long time period (e.g. one year for car insurance), so there exists data sparsity and cold start problem. Researchers try to solve the problem by recommendation systems (RS) \cite{DBLP:conf/sac/RokachSSCS13,DBLP:conf/recsys/QaziFMF17,DBLP:conf/icmcs/LiuZKZZ019}, however, these methods directly apply traditional RS model to insurance domain, neglecting item complexity and data sparsity.

PingAn Jinguanjia (PAJGJ) is one of the most popular comprehensive applications (APP) in China. In addition to traditional e-commerce products (defined as nonfinancial products in this paper), e.g. household supplies, it also provides financial products like insurance products, investment services. Here we focus on recommending insurance products. Traditional RS, like collaborate filtering (CF) could not perform effectively in insurance domain for its particular characteristics. To get more accurate recommendation, our company tries to use side information from PAJGJ (interaction behaviors from nonfinancial domain), but to little avail. 

Cross-domain recommendation (CDR) \cite{DBLP:conf/ijcai/ManSJC17,DBLP:conf/sigir/MaRLCMR19,DBLP:conf/cikm/KangHLY19}, employing data from multiple domains, is one of the promising ways to solve data sparsity and cold start problem. Generally, CDR can be categorized into two categories. One is interested in improving the overall performance in target domain by aggregating knowledge between two domains \cite{DBLP:conf/sigir/MaRLCMR19}. The other one aims at infering the preferences of cold start users based on their preferences observed in other domains  \cite{DBLP:conf/ijcai/ManSJC17,DBLP:conf/cikm/KangHLY19}. These methods assume that there exists overlap in information between users and/or items across different domains, and train a mapping function from source domain into target domain. Unfortunately, we could not apply CDR methods into insurance and nonfinancial domain directly for its properties. 

Based on the observations, we propose a novel framework called a Deep Cross-Domain Insurance Recommendation System (DCDIR) for cold start users. Specifically, we first try to learn more effective user and item latent features in both source and target domains. In target domain, given the complexity of insurance products, we design a meta-path based method over the knowledge graph we constructed. In source domain, we employ gated recurrent unit (GRU) to model users' dynamic interests. After obtaining the latent features of the overlapping users, a feature mapping function between the two domains is learned by multi-layer perceptron (MLP).
 
In summary, our contributions in this paper are as follows:
\begin{itemize}
\item To the best of our knowledge, this is the first work to utilize cross-domain mechanism to give personalized recommendations for cold start users in insurance domain.
\item For the complexity of insurance products, we design a meta-path based method to learn more effective latent user and item features, revealing reasons behind recommendations.
\item We conduct experiments on our company's scenarios, the results prove the efficacy of DCDIR over several baselines.
\end{itemize}

\section{Problem Formulation}

Let $\mathcal{U}=\{u_{1},u_{2},\ldots,u_{m}\}$ denote overlapping users between nonfinancial domain (source domain) $\mathcal{D}^{s}$ and insurance domain (target domain) $\mathcal{D}^{t}$. If a user only appears in one domain, he/she is a cold start user in the other domain. The user-item interaction matrices are denoted as $\boldsymbol{Y}^{s}\in\mathbb{R}^{m\times s}$ and $\boldsymbol{Y}^{t}\in\mathbb{R}^{m\times t}$, which are defined according to users’ implicit feedbacks. We additionally use $\mathcal{NI}_{u}^{s}$ and $\mathcal{NI}_{u}^{t}$ for the sequences of items that user $u$ has interacted with. We also have a insurance knowledge graph (ISKG) $\mathcal{G}$, which consists of multiple entity types (i.e. Product, Feature, Need) and many entity-relation-entity triples $(h,r,t)$. For example, (\emph{travel accident insurance, insurance.product-insurance.type, accident insurance}) states the type of ``travel accident insurance'' is accident insurance. Given rating matrices and ISKG, our goal is to learn the mapping function from nonfinancial domain to insurance domain, which can help us deal with cold start users.

\section {DCDIR}

\begin{figure}[!h]
\setlength{\abovecaptionskip}{0.cm}
\setlength{\belowcaptionskip}{0.cm}
  \centering
  \includegraphics[scale=0.44]{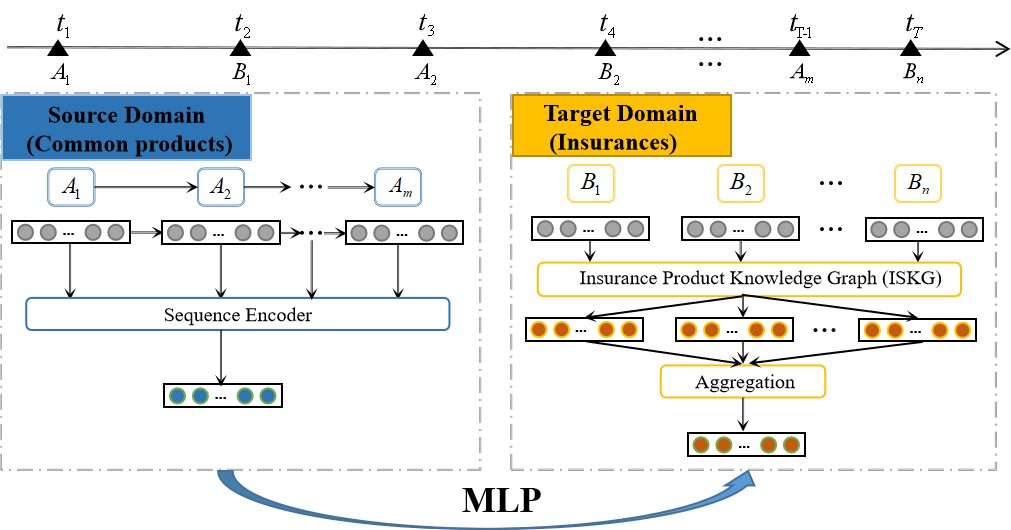}
  \caption{The Framework of DCDIR}
  \label{SIGIR2020_FRAMEWORK}
\end{figure}

To provide recommendations to cold start users, we propose DCDIR. As shown in Figure \ref{SIGIR2020_FRAMEWORK}, DCDIR contains three main parts: learning user latent features in two domain, mapping of user latent features.

\subsection{Latent Feature in Target Domain}

\begin{figure}[!h]
  \centering
  \includegraphics[scale=0.25]{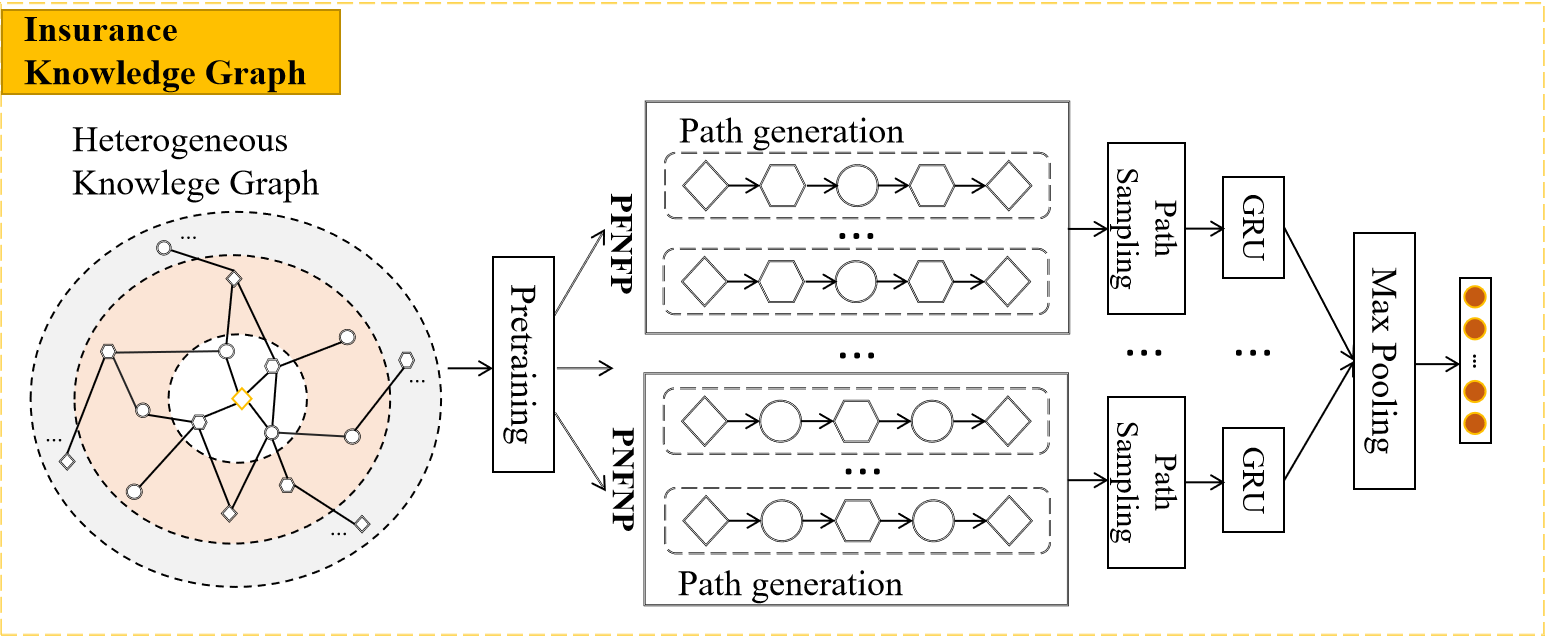}
  \caption{Meta-Path based ISKG Module}
  \label{SIGIR2020_KG_path}
\end{figure}

As mentioned above, the complexity of insurance products is typically non-trivial, understanding the items may require a considerable cognitive overload \cite{DBLP:conf/sac/RokachSSCS13}. To help users better understand insurance products, we design a meta-path based  method. Figure \ref{SIGIR2020_KG_path} shows the framework, we first pretrain KG by TransD \cite{DBLP:conf/acl/JiHXL015}, and get entity and relation embeddings, which are denoted by $\boldsymbol{e}$, $\boldsymbol{r} \in\mathbb{R}^{d}$. Then, we generate meta-paths connecting user's interacted items and target item. To select high-quality meta-paths, we properly design a score function. Finally, we use GRU to model each meta-path and employ max-pooling to aggregate these selected paths.

\subsubsection{Path Generation} 

The triples in KG describe relational properties of items, which constitute several paths between the user's interacted items and target item. For a given user $u$, we formally define the path from $i\in\mathcal{NI}_{u}^{t}$ to target item $v$ as a sequence of entities and relations: $p_{e_{1},e_{L}}=[e_{1}\xrightarrow{r_{1}} e_{2}\xrightarrow{r_{2}}\ldots \xrightarrow{r_{L-1}}e_{L}]$, where $e_{1}=i\in\mathcal{NI}_{u}^{t}$, $e_{L}=v$, $(e_{l},r_{l},e_{l+1})$ is the $l$-th triple in $p_{e_{1},e_{L}}$, and $L$ denotes the number of triples in the path. We use $\mathcal {S}_{u}=\left\{p_{e_{1},e_{L}}|e_{1}\in\mathcal{NI}_{u}^{t}\right\}$ to denote all generated paths of $u$. From the construction of ISKG, we know that relation $r_{l-1}$ and entity $e_{l}$ have similar semantics, so the embedding of $p_{e_{1},e_{L}}$ is denoted as $[\boldsymbol{e}_{1}, \boldsymbol{e}_{2},\ldots, \boldsymbol{e}_{L}]$. 
Long meta-paths are likely to introduce noisy semantics \cite{DBLP:journals/pvldb/SunHYYW11}, we properly design two meta-paths based on our scenario, where we fix entity type and path length. They are represented as $(P,F,N,F,P)$ and $(P,N,F,N,P)$, where $P$, $N$ and $F$ denote ``Product'' , ``Need'' and ``Feature''. Here are two examples.
\begin{eqnarray*}
&(P,F,N,F,P)\;\;GEF\rightarrow\text{compensate critical illness}\rightarrow&\\
&\text{insure critical illness}\rightarrow\text{high premium}\rightarrow AXBFB&\\
&(P,N,F,N,P)\;\;ESB \rightarrow\text{insure medical treatment}\rightarrow&\\
&\text{high level assurance}\rightarrow\text{insure accident}\rightarrow BWRWX&
\end{eqnarray*}
where $GEF$ is critical illness insurance, $EBS$ is health insurance, $AXBFB$ and $BWRWX$ are accident insurances.

\subsubsection{Sampling Top-K High Quality Path Instances}
\label{select}

There still so many meta-paths, even though we have fixed path structure. Some of the paths bring much more noises than useful signals, so we use $top$-$K$ sampling module to select $K$ useful paths. Specifically, for a given path $p_{e_{1},e_{L}}=[e_{1}, e_{2},\ldots, e_{L}]$, we define a score function:
\begin{eqnarray}
\setlength{\abovedisplayskip}{0pt}
\setlength{\belowdisplayskip}{0pt}
\label{SIGIR2020_pathscore}
s_{e_{1},e_{L}}=\text{softmax}(\frac{P}{|\mathcal{NI}_{u}^{t}|})+\frac{\boldsymbol{e}_{L}^{T}}{\|\boldsymbol{e}_{L}\|}\sum_{i=1}^{L-1}\frac{\boldsymbol{e}_{i}}{\|\boldsymbol{e}_{i}\|},
\end{eqnarray}
where $P$ is $e_{1}$'s position in $\mathcal{NI}_{u}^{t}$. The first part of \eqref{SIGIR2020_pathscore} is to measure interaction time, since more recent items in a sequence have a larger impact on users’ next actions. The second part is to measure the similarity between the path and the target item. For a user, we select top-K paths with high score, which are denoted by a set  $\mathcal{S}_{u}^{top-K}$, $K$ is a given parameter.

\subsubsection{Path Embedding and User Feature Representation} 

A path instance is a node entity sequence, to embed such sequence into a low-dimensional vector, we take GRU \cite{DBLP:journals/corr/HidasiKBT15}.
The formulations are:
\begin{eqnarray}
\setlength{\abovedisplayskip}{0pt}
\setlength{\belowdisplayskip}{0pt}
\begin{aligned}
\label{SIGIR20_GRU}
\boldsymbol{x}_{n}=\;&\sigma(\boldsymbol{W}_{x}\boldsymbol{e}_{n}+\boldsymbol{U}_{x}\boldsymbol{h}_{n-1}+\boldsymbol{b}_{x})\\
\boldsymbol{r}_{n}=\;&\sigma(\boldsymbol{W}_{r}\boldsymbol{e}_{n}+\boldsymbol{U}_{r}\boldsymbol{h}_{n-1}+\boldsymbol{b}_{r})\\
\widetilde{\boldsymbol{h}}_{n} = \;&\tanh(\boldsymbol{W}_{h}\boldsymbol{e}_{n}+\boldsymbol{r}_{n}\circ\boldsymbol{U}_{h}\boldsymbol{h}_{n-1}+\boldsymbol{b}_{h})\\
\boldsymbol{h}_{n} = \;&(\boldsymbol{1}-\boldsymbol{x}_{n})\circ \boldsymbol{h}_{n-1}+\boldsymbol{x}_{n}\circ\widetilde{\boldsymbol{h}}_{n},
\end{aligned}
\end{eqnarray}
where $\sigma$ is sigmoid function,
$\circ$ is element-wise product, $\boldsymbol{W}_{x}$, $\boldsymbol{W}_{r}$, $\boldsymbol{W}_{h} \in\mathbb{R}^{n_{H}\times d}$, $\boldsymbol{U}_{x}$, $\boldsymbol{U}_{r}$, $\boldsymbol{U}_{h}\in\mathbb{R}^{n_{H}\times n_{H}}$, $n_{H}=d$ is hidden size. Let $\boldsymbol{p}_{e_{1},e_{L}}=\boldsymbol{h}_{n}$, and apply max-pooling, i.e.:
\begin{eqnarray*}
\textbf{u}^{t} = \text{max-pooling}\left\{\boldsymbol{p}_{e_{1},e_{L}}|p_{e_{1},e_{L}}\in\mathcal{S}_{u}^{top-K}\right\}.
\end{eqnarray*}

\subsection{Latent Feature in Source Domain}

In our APP, each item $i$ in nonfinancial domain is associated with a description $c_{i}$. In order to learn more effective latent features, we employ word2vec \cite{DBLP:conf/nips/MikolovSCCD13}.
Suppose there are $n$ words in $i$'s content $c_{i}$. We utilize word2vec to obtain word vectors, which are represented as $\{\boldsymbol{w}^{i}_{k}\}_{k=1}^{n}$. Then we get the final item embedding by:
\begin{eqnarray*}
\boldsymbol{i}=\text{max-pooling}(\text{concat}(\{\boldsymbol{w}^{i}_{k}\}_{k=1}^{n})).
\end{eqnarray*}
To model user latent feature $\boldsymbol{u}^{s}$, we employ GRU over $\mathcal{NI}_{u}^{s}$, and let $\boldsymbol{u}^{s}=\boldsymbol{h}^{\text{GRU}}_{n}\left(\mathcal{NI}_{u}^{s}\right)$, the equation is replacing $\boldsymbol{e}_{n}$ by $\boldsymbol{i}_{n}$ in eq. \eqref{SIGIR20_GRU}.

\subsection{Mapping Function Between Two Domains}

We employ MLP \cite{DBLP:conf/ijcai/ManSJC17} to learn mapping function between two domains, taking $\boldsymbol{u}^{s}$ as input and $\boldsymbol{u}^{t}$ as output. The loss function is:
\begin{eqnarray*}
\setlength{\abovedisplayskip}{2pt}
\setlength{\belowdisplayskip}{0pt}
\mathcal{L}_{\text{cross}} = \sum_{u\in\mathcal{U}}\|f_{\text{mlp}}(\boldsymbol{u}^{s})-\boldsymbol{u}^{t}\|_{2}.
\end{eqnarray*}
\subsection{Training}

In the training process, loss functions for each part is added together for joint optimization. The overall loss function is:
\begin{eqnarray*}
\mathcal{L} = \mathcal{L}_{\text{cross}}+\mathcal{L}_{T}+\mathcal{L}_{S}.
\end{eqnarray*}
where $\mathcal {L}_{T}$ and $\mathcal {L}_{S}$ are recommendation loss in target and source domain, respectively. Take the target domain as an example, 
\begin{eqnarray*}
\mathcal{L}_{T}=\sum_{(u,v)\in Y^{t}}-\left(y_{uv}\log\hat{y}_{uv}+(1-y_{uv})\log(1-\hat{y}_{uv})\right),
\end{eqnarray*}
where $
\hat{y}_{uv}=\sigma(f(\boldsymbol{u}^{t},\boldsymbol{v}^{t}))
$,
$\sigma(\cdot)$ is sigmoid function, $f$ is a ranking function which can be a dot-product or a deep neural network.

\subsection{Cross-Domain Recommendation}

In this paper, we assume cold start users have interactions in nonfinancial domain, but no interactions in insurance domain. After learning the latent features in nonfinancial domain $\boldsymbol{u}^{s}$, we can get the corresponding mapping latent features $\hat{\boldsymbol{u}}^{t}=f_{\text{mlp}}(\boldsymbol{u}^{s})$. Based on learned $\hat{\boldsymbol{u}}^{t}$, we can make recommendations to cold start users.

\section{Experiment}

We conduct extensive experiments to answer the following questions: RQ1: How does DCDIR model perform compared with baselines in terms of NDCG and Recall@3? RQ2: Can DCDIR alleviate the data sparsity problem? RQ3: How does path-based ISKG module affect the performance of DCDIR for cold start users?

\begin{table}[!t]
\setlength{\abovecaptionskip}{0.cm}
\setlength{\belowcaptionskip}{0.cm}
\centering
\footnotesize{
\caption{Statistics of the JGJISNF dataset.}
\label{dataset}
\begin{tabular}{c|c|c|c}
\toprule[1pt]
\multicolumn{2}{c|}{IS-domain (Target domain)}&\multicolumn{2}{c}{NF-domain (Source domain)	}\\
\hline
$\#$Items				&	42		&	$\#$Items				&	3,836		\\
$\#$Interactions				&	300,000		&	$\#$Interactions				&	600,000		\\
$\#$KG relations				&	7		&					&		\\
$\#$KG enitities				&	77		&					&			\\
$\#$KG triples				&	282		&					&			\\
\hline
\multicolumn{2}{c|}{$\#$Overlapped-users}&\multicolumn{2}{c}{21,016}\\
\multicolumn{2}{c|}{$\#$Training-sequences}&\multicolumn{2}{c}{12,437}\\
\multicolumn{2}{c|}{$\#$Test-sequences}&\multicolumn{2}{c}{4,218}\\
\multicolumn{2}{c|}{$\#$Validation-sequences}&\multicolumn{2}{c}{4,298}\\
\bottomrule[1.0pt]
\end{tabular}
}
\end{table}

\subsection{Experimental Settings}
\label{es}

\textbf{Datasets.} There is no publicly available dataset for CDR-ISNF (cross-domain recommendation for insurances and nonfinancial products). To demonstrate the overall effectiveness of the proposed DCDIR model, we build and release a sub-dataset (named JGJISNF) from a comprehensive e-commerce dataset that contains about 20 million users pursue logs from June 1st 2018 to May 31th 2019. The pursue logs are collected on IS-domain and NF-domain from a well-known e-commerce platform PAJGJ. The IS-domain contains short-term insurances (periods is less than 1 year, e.g., including illness insurances, accident insurances, etc.) interactions . The NF-domain contains user logs of non-financial products (daily necessities products, e.g.,clothes, skincare products, fruits, electronics products, etc). In the two domains, we gather chronological user behaviors, user profiles and detailed product descriptions. Due to the complexity of insurance products, we construct a knowledge graph of insurance products based on their own information.

\begin{table*}[!t]
\setlength{\abovecaptionskip}{0.cm}
\setlength{\belowcaptionskip}{0.cm}
\centering
\footnotesize{
\caption{Performance comparison in Recall@3 and NDCG. The best baseline except DCDIR is bolded. Numbers in ``()'' represent the percentage of three variants' performance at $\eta$=10$\%$ compared with their best performance in other sparsity level.}
\label{baseline}
\begin{tabular}{c|c|c|c|c|c|c|c|c}
\toprule[1pt]
$\eta$&\multicolumn{2}{c|}{10$\%$}&\multicolumn{2}{c|}{20$\%$}&\multicolumn{2}{c|}{50$\%$}&\multicolumn{2}{c}{100$\%$}\\
\hline
Method	&	NDCG	&	Recall@3	&	NDCG	&	Recall@3	&	NDCG	&	Recall@3	&	NDCG	&	Recall@3	\\
\hline
BPR	&	0.27011	&	0.06418	&	0.27105	&	0.06518	&	0.27133	&	0.06451	&	0.27325	&	0.07124	\\
\hline
GRU4REC	&	0.23923	&	0.02143	&	0.25964	&	0.07768	&	0.30725	&	0.09611	&	0.30623	&	0.08602	\\
\hline
EMCDR-BPR	&	0.27343	&	0.07291	&	0.27342	&	0.07291	&	0.27342	&	0.07325	&	0.27347	&	0.07325	\\
\hline
EMCDR-GRU	&	0.26775	&	0.11794	&	0.26801	&	0.11794	&	0.29056	&	0.11996	&	0.31288	&	0.12298	\\
\hline
DCDIR-V1	&	0.34781(-4.66$\%$)	&	0.17321(-6.28$\%$)	&	0.35196	&	0.18016	&	0.35653	&	0.18078	&	0.36481	&	0.18481	\\
\hline
DCDIR-V2	&	{\bf 0.36278}(-13.05$\%$)	&	{\bf 0.19159}(-27.60$\%$)	&	{\bf 0.37021}	&	{\bf 0.19388}	&	{\bf 0.40273}	&	{\bf 0.24504}	&	{\bf 0.40925}	&	{\bf 0.26461}	\\
\hline
DCDIR	&	0.39394(-3.95$\%$)	&	0.25185(-5.31$\%$)	&	0.39741	&	0.25227	&	0.40773	&	0.26268	&	0.41016	&	0.26597	\\
\hline
DCDIR vs. best	&	8.59$\%$	&	26.23$\%$	&	7.35$\%$	&	24.96$\%$	&	1.24$\%$	&	7.20$\%$	&	0.22$\%$	&	0.51$\%$	\\
\bottomrule[1.0pt]
\end{tabular}
}
\end{table*}

\textbf{Comparative Models and Metrics.} 
We compare DCDIR with four baselines and two variants of DCDIR. The baselines can be categorized into single-domain group (BPR\cite{DBLP:conf/uai/RendleFGS09} and GRU4REC \cite{DBLP:journals/corr/HidasiKBT15}) and cross-domain group (EMCDR-BPR \cite{DBLP:conf/ijcai/ManSJC17},EMCDR-GRU, DCDIR, DCDIR-V1 and DCDIR-V2). The first group is to validate the usefulness of CDR models, and the second group is for demonstrating the advantage of path-based method. DCDIR leverage path-based method to deal with insurance products' complex knowledge graph, while DCDIR-V1 and DCDIR-V2 use only simple products' attributes and KGE method (2-hop entity aggregation among ISKG), respectively. 

We evaluate all models in terms of Recall@N (N=3) and NDCG. We adopt a common and widely used strategy to avoid heavy computation on evaluating all user-item pairs {\cite{DBLP:conf/www/HeLZNHC17,DBLP:conf/aaai/WangWX00C19,DBLP:conf/kdd/Wang00LC19}}. For each user $u$, we randomly sample negative items that don't appear in the training set and rank them with the single ground-truth item.

\textbf{Parameter Setting.} We randomly select 30$\%$ of the total overlapped users and remove their information in the target domain as cold start users for evaluating the performance (i.e., test users). To study the performance of DCDIR with respect to the number of overlapped users, we restrict the number of the overlapped users similarly to the real-world distribution. We build four training sets with a certain fraction $\eta\in\{10\%, 20\%, 50\%, 100\%\}$ of overlapped users who do not belong to the test users. These settings are chosen with grid search on the validation set. Item embedding size and GRU hidden state size are set to 50. We use dropout with drop ratio $p = 0.8$. For the parameters in Section \ref{select} (path-based method section), we try different settings, the analysis of which can be found in Section \ref{rq3}.For the hyper-parameters of the Adam optimizer,we set the learning rate $\alpha$= 0.001. To speed up the training and converge quickly, batch size is set to 32. We test the model performance on the validation set for every epoch.

\subsection{Performance Comparison (RQ1 and RQ2)}
To answer RQ1 and RQ2, three variants of DCDIR are compared with four state-of-the-art models with different densities. Table \ref{baseline} shows the performance comparison. Overall, benefiting from the proposed insurance products' KG path-based representations and source domain information, DCDIR beats all comparative methods, and achieves the range of 0.22$\%$-8.59$\%$ and 0.51$\%$-26.23$\%$ improvements over the best comparative model in Recall@3 and NDCG under all levels of data sparsity, respectively. These experiments reveal a number of interesting discoveries: (1) All cross-domain methods yield better performances than single-domain methods with mixture of target and source domain data , demonstrating the importance of cross-domain module; (2) Owing to the capability of using insurance products’ knowledge, three variants of DCDIR (DCDIR, DCDIR-V1 and DCDIR-V2) defeat other comparative methods; (3) It also demonstrates that DCDIR achieves more improvements in a sparser dataset than in a denser one. It is validated that, compared to comparative approaches, DCDIR can better diminish the negative impacts of the data sparsity issue. We also conduct experiments to compare DCDIR with DCDIR-V1 and DCDIR-V2 (definition refer to \ref{es} comparative models). Numbers in “( )” shows the performance of DCDIR-V2 using KGE method declines sharply in terms of Recall@3 (-13.05$\%$) and NDCG (-27.60$\%$) when using a sparser dataset, while DICIR-V1 cannot outperform DICDIR in all levels of sparsity. This shows that, DCDIR can get more stable and better performance with limited data.

\begin{table}[!h]
\setlength{\abovecaptionskip}{0.cm}
\setlength{\belowcaptionskip}{0.cm}
\centering
\footnotesize{
\caption{Performance comparison in Recall@3 and NDCG under a sparse setting ($\eta$=10$\%$) with changing path number and choosing path strategy.}
\label{ques3}
\begin{tabular}{c|c|c|c}
\toprule[1pt]
\multicolumn{2}{c|}{ISKG module}&\multicolumn{2}{c}{Metrics}\\
\hline
parameter	&	value	&	NDCG	&	Recall@3	\\
\hline
\multirow{3}{*}{path$\_$num}	&	10	&	0.36611	&	0.18207	\\
\cline{2-4}
	&	20	&	0.39394	&	0.25185	\\
	\cline{2-4}
	&	30	&	0.38435	&	0.18541	\\
	\hline
\multirow{2}{*}{path$\_$strategy}	&	`topk'	&	0.39394	&	0.25185	\\
\cline{2-4}
	&	`random'	&	0.34624	&	0.16065	\\
\bottomrule[1.0pt]
\end{tabular}
}
\end{table}

\subsection{The impact of meta-path based ISKG module to cold start users (RQ 3)}
\label{rq3}
The cold start problem is one of the major challenges for RS. It is necessary to study if our designed  meta-path based ISKG module can deal with cold start users problem in an effective way. Therefore, we compare DCDIR with different parameters' value, the number of path selected and strategy of choosing high-quality paths, in an extremely sparse dataset with $\eta$=10$\%$, where the segmentation of training, testing and validation dataset as introduced above. Table \ref{ques3} indicates that, suffering from the cold start problem, DCDIR's best parameters in ISKG module are path number as 20 and choosing path strategy is our designed top K method in terms of Recall@3 and NCDG. Specifically, path strategy can effect the performance of DCDIR significantly with a large improvement in Recall@3 and NCDG, respectively. Top K strategy optimizes the choice of high-quality insurance products' KG paths, which both leverage rich and complicated information and interference information. Therefroe, DCDIR can better handle cold start users.

\section{Conclusion}

To deal with insurance product complexity and cold start problem, we propose DCDIR for cold start users. Specifically, we first learn more effective user and item latent features in two domains. In target domain, given the complexity of insurance products, we design a meta-path based method over insurance product knowledge graph, which can provide interpretable recommendations to users. In source domain, we employ GRU to model users' dynamic interests.  Then we learn a feature mapping function by multi-layer perceptions . We apply DCDIR on our company’s dataset, and show DCDIR significantly outperforms the state-of-the-art solutions.

\bibliographystyle{ACM-Reference-Format}
\bibliography{REF_SIGIR_2020}
\end{document}